\documentclass[runningheads]{llncs}
\usepackage[T1]{fontenc}
\usepackage{booktabs}
\usepackage{graphicx}
\usepackage{tikz}
\usetikzlibrary{arrows}
\usepackage{verbatim}
\usepackage{array,multirow}
\usepackage{hyphenat}
\usetikzlibrary{shapes}
\usetikzlibrary{calc}
\begin{document}
\title{Automatic Breast Lesion Classification by Joint Neural Analysis of Mammography and Ultrasound}
\titlerunning{Multimodal Breast Cancer Classification}
%
\author{Gavriel Habib\inst{1} \and
Nahum Kiryati\inst{2} \and Miri Sklair-Levy\inst{3} \and
Anat Shalmon\inst{3} \and Osnat Halshtok Neiman\inst{3} \and Renata Faermann Weidenfeld\inst{3} \and Yael Yagil\inst{3} \and Eli Konen\inst{3} \and Arnaldo Mayer\inst{3}}


\authorrunning{Habib et al.}
%

\institute{School of Electrical Engineering, Tel-Aviv University, Tel Aviv-Yafo, Israel\\ 
\email{gavrielhabib@mail.tau.ac.il}\\ \and
The Manuel and Raquel Klachky Chair of Image Processing,
School of Electrical Engineering, Tel-Aviv University, Tel Aviv-Yafo, Israel \and Diagnostic Imaging, Sheba Medical Center, affiliated to the Sackler School of Medicine, Tel-Aviv University, Israel}

\maketitle              
\begin{abstract}
Mammography and ultrasound are extensively used by radiologists as complementary modalities to achieve better performance in breast cancer diagnosis. However, existing computer-aided diagnosis (CAD) systems for the breast are generally based on a single modality. In this work, we propose a deep-learning based method for classifying breast cancer lesions from their respective mammography and ultrasound images. We present various approaches and show a consistent improvement in performance when utilizing both modalities. The proposed approach is based on a GoogleNet architecture, fine-tuned  for our data in two training steps. First, a distinct neural network is trained separately for each  modality, generating high-level features. Then, the aggregated features originating from each modality are used to train a multimodal network to provide the final classification. In quantitative experiments, the proposed approach achieves an AUC of 0.94, outperforming state-of-the-art models trained over a single modality. Moreover, it performs similarly to an average radiologist, surpassing two out of four radiologists participating in a reader study. The promising results suggest that the proposed method may become a valuable decision support tool for breast radiologists.

\end{abstract}
\section{Introduction}
\label{sec:intro}

Breast cancer is the second most common type of cancer among American women after skin cancer. According to the American Cancer Society estimations, 268,600 invasive breast cancer cases have been diagnosed in 2019, leading to 41,760 deaths. However, early detection may save lives as it enables better treatment options. 

Mammography-based screening is the most widely used approach for breast cancer detection,  with proven mortality reduction and early disease treatment benefits~\cite{ref_mammo_advantages}. However, it suffers from poor lesion visibility in dense breasts~\cite{mammo_overview}. To improve sensitivity in dense breasts, contrast-enhanced spectral mammography (CESM) has been developed. CESM is based on the subtraction of low and high energy images, acquired following the injection of a contrast agent~\cite{CESM}. Although CESM reaches MRI levels of lesion visibility for dense breasts~\cite{CESM_MRI}, the technique is still in the early adoption phase.  

Ultrasound imaging has proven to be a valuable tool in dense breasts, increasing cancer detection sensitivity by 17\%  \cite{US_dense_breasts}. Nevertheless, breast ultrasound may miss solid tumors that are easily detected with mammography. Devolli-Disha et al. \cite{mammo_US_specificity_sensitivity} showed that ultrasound had a higher sensitivity (69.2\%) than mammography (15.4\%) in women younger than 40 years, whereas mammography  (78.7\%) beats ultrasound (63.9\%) in women older than 60 years. Due to its benefits and disadvantages, radiologists suggest using breast ultrasound as a complementary screening test to mammography~\cite{radiologists_combined_1,radiologists_combined_2}. 

\begin{figure}[t]  
\centering
\includegraphics[width=0.25\textwidth]{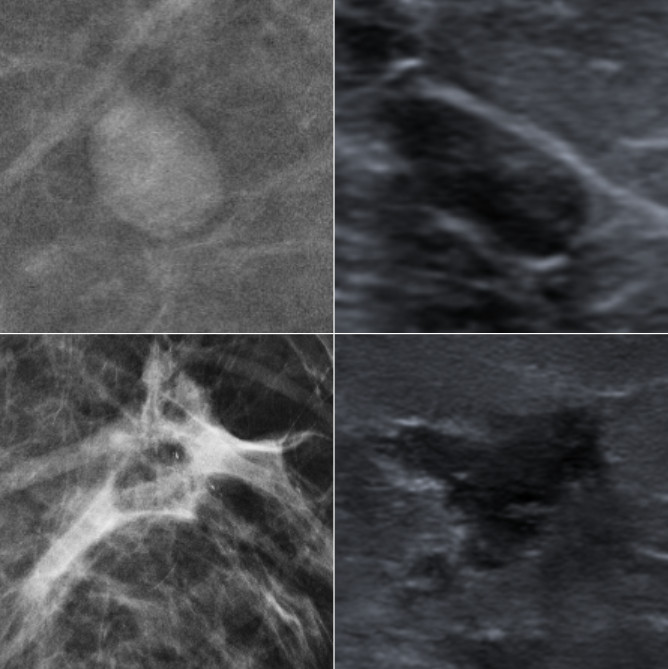}
\caption{Benign (Top) and Malignant (Bottom) lesions from our dataset. Malignant lesions tend to have less strict boundaries in both mammography (left) and ultrasound (right) screenings.}
\label{fig:benign_malignant_samples}
\end{figure}

Classification of breast lesions is a challenging task for the radiologist. Malignant and benign lesions can be differentiated by their shape, boundary and texture. For example, malignant lesions may have irregular and not well defined boundaries as they have the ability to spread (see Figure~\ref{fig:benign_malignant_samples}). Nevertheless, in many cases radiologists cannot classify the lesion and the patient is referred for a biopsy which is a stressful and expensive process. Given that 65\%-85\% of the biopsies turns out to be benign~\cite{biopsies_rate}, there is a clear need for tools that will help radiologists reduce benign biopsies.

\section{Related Work}
\label{sec:related_work}

In recent years, deep learning techniques have been providing significant improvements in various medical imaging tasks, such as tumor detection and classification, image denoising and registration. In the field of breast cancer classification, existing methods are based mainly on mammograms~\cite{mammo_transfer_learning,NYU_mammo,mammo_classification_attention,mammo_GAN}, ultrasound~\cite{us_semi_supervised,US_classification}, MRI~\cite{MRI_classification_1,MRI_classification_2} or histopathology images~\cite{HP_classification}.

To deal with the limited amount of data, Chougrada et al.~\cite{mammo_transfer_learning} used transfer learning over ImageNet and achieved state-of-the-art results over public mammography datasets. Cheng et al.~\cite{us_semi_supervised} performed a semi-supervised learning approach over a large breast ultrasound dataset with only few annotated images. Wu et al.~\cite{mammo_GAN} synthesized mammogram lesions using class-conditional GAN and used them as additional training data instead of basic augmentations.

Emphasizing the importance of lesions' context, Wu et al.~\cite{NYU_mammo} trained a deep multi-view CNN over a large private mammogram dataset. They used a breast-level model to create heatmaps that represent suspected areas, and a patch-level model to locally predict the presence of malignant or benign findings. Shen et al.~\cite{mammo_classification_attention} combined coarse and fine details using an attention mechanism to select informative patches for classification.

Common breast imaging modalities were also combined with additional data from other domains. Byra et al.~\cite{us_nakagami} used the Nakagami parameter maps created from breast ultrasound images to train a CNN from scratch. Perek et al.~\cite{shaked_CESM} integrated CESM images with features of BIRADS~\cite{BIRADS}, a textual radiological lexicon for breast lesions, as inputs to a classifier.

Most of previous studies utilized only a single modality, while some combined different types of breast images. Hadad et al.~\cite{IBM_MRI_MG} classified MRI breast lesions using fine tuning of a network pre-trained on mammography images instead of natural images. Regarding mammography with ultrasound, Cong et al.~\cite{MG_US_ensemble_classification} separately trained three base classifiers (SVM, KNN and Naive Bayes) for each modality, integrated some of them by a selective ensemble method and obtained the final prediction by majority vote. Shaikh et al.~\cite{LUPI_based_classifier} proposed a learning-using-privileged-information approach, i.e. utilizing both modalities for training, but avoiding one during test time. These papers suggested the potential of cross-modal learning.

In this paper, we propose a novel deep-learning  method for the classification of breast lesion, using both mammography and ultrasound images of the lesion. To the best of our knowledge, it is the first reported attempt to combine these very different imaging modalities by fusing high-level perceptual representations for lesion classification.  We use a unique dataset consisting of matched mammography and ultrasound lesions, acquired at our institution. The proposed methods are evaluated  using a leave one out scheme, demonstrating significant improvement in AUC (area under curve) when  features extracted from both modalities are combined into a single multi-modality classifier, in comparison to single modality classification using only mammography or ultrasound.

\section{Method}

\subsection{Dataset}
\begin{figure}[!t]
\centering
\includegraphics[width=0.3\textwidth]{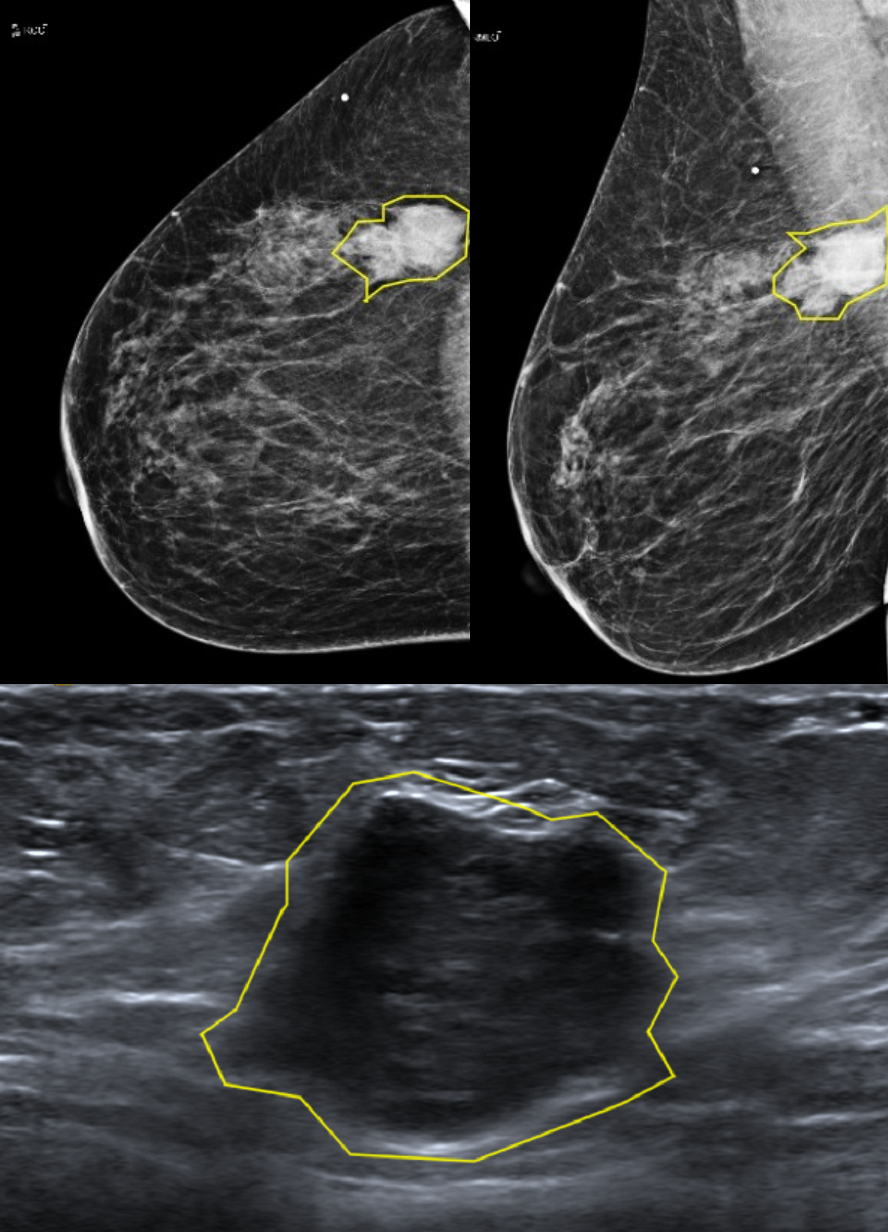}
\caption{Matched malignant lesion contouring in both modalities.}
\label{fig:contours}
\end{figure}

Although combining mammography and ultrasound imaging for breast cancer screenings is a common practice, to the best of our knowledge there are no public datasets containing corresponding lesions from both modalities. Therefore, we created our own retrospective dataset of 153 biopsy-proven lesions, consisting of 73 malignant and 80 benign cases. For each lesion, corresponding mammography and ultrasound images were contoured by an expert breast radiologist, with a biopsy proven labelling. Figure~\ref{fig:contours} demonstrates a sample from the dataset.

\subsection{Model Architecture}
\subsubsection{Single modality networks} Two convolutional neural networks (CNNs), one for each modality, were trained to tell apart malignant and benign lesions. The contoured lesions were cropped into image patches and submitted to geometric transformations (translation, rotation, flipping) to augment the dataset and generate additional inputs.

\begin {figure}[!t]
\centering
\resizebox {0.65\textwidth} {!} { %
\centering
\begin{tikzpicture}

\pgfmathsetmacro{\cubex}{0.1}
\pgfmathsetmacro{\cubey}{4}
\pgfmathsetmacro{\cubez}{5}

\pgfmathsetmacro{\multx}{1.5}
\pgfmathsetmacro{\divyz}{2}

\definecolor{convfill}{RGB}{255,255,255}
\definecolor{convline}{RGB}{0,0,0}

\definecolor{maxline}{RGB}{255,0,0}
\definecolor{maxfill}{RGB}{255,255,255}

\definecolor{fcfill}{RGB}{255,255,255}
\definecolor{fcline}{RGB}{6,185,251}
\definecolor{fcdescmammo}{RGB}{20,84,179}

\definecolor{softmaxfill}{RGB}{255,255,255}
\definecolor{softmaxline}{RGB}{191,115,16}

\pgfmathsetmacro{\x}{0}
\pgfmathsetmacro{\offset}{0.08}

\pgfmathsetmacro{\xx}{0.5}

\node[inner sep=0pt,cm={0.37 ,0.5 ,0 ,1  ,(0 cm,0 cm)}] (image) at (-1.6,0)
    {\includegraphics[width=4 cm]{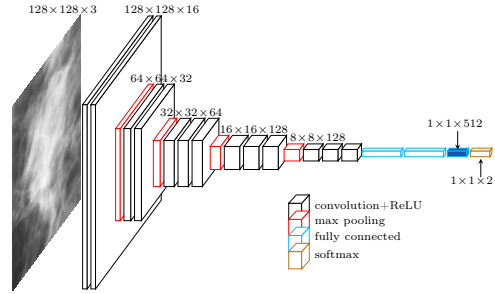}};
  
\node[text width=3cm] at (\x+\xx/2-\xx-1.2,\cubey/2+0.12,\cubez/2-\cubez) {\scriptsize $128 \! \times \! 128 \! \times \! 3$};

\draw[convline,fill=convfill] (\x+\xx/2,\cubey/2,\cubez/2) -- ++(-\cubex,0,0) -- ++(0,-\cubey,0) -- ++(\cubex,0,0) -- cycle;
\draw[convline,fill=convfill] (\x+\xx/2,\cubey/2,\cubez/2) -- ++(0-\xx,0,-\cubez) -- ++(0,-\cubey,0) -- ++(0+\xx,0,\cubez) -- cycle;
\draw[convline,fill=convfill] (\x+\xx/2,\cubey/2,\cubez/2) -- ++(-\cubex,0,0) -- ++(0-\xx,0,-\cubez) -- ++(\cubex,0,0) -- cycle;

\pgfmathsetmacro{\x}{\x + \cubex + \offset}
\draw[convline,fill=convfill] (\x+\xx/2,\cubey/2,\cubez/2) -- ++(-\cubex,0,0) -- ++(0,-\cubey,0) -- ++(\cubex,0,0) -- cycle;
\draw[convline,fill=convfill] (\x+\xx/2,\cubey/2,\cubez/2) -- ++(0-\xx,0,-\cubez) -- ++(0,-\cubey,0) -- ++(0+\xx,0,\cubez) -- cycle;
\draw[convline,fill=convfill] (\x+\xx/2,\cubey/2,\cubez/2) -- ++(-\cubex,0,0) -- ++(0-\xx,0,-\cubez) -- ++(\cubex,0,0) -- cycle;

\node[text width=3cm] at (\x+\xx/2-\xx+0.7,\cubey/2+0.12,\cubez/2-\cubez) {\scriptsize $128 \! \times \! 128 \! \times \! 16$};

\pgfmathsetmacro{\cubey}{\cubey/\divyz}
\pgfmathsetmacro{\cubez}{\cubez/\divyz}
\pgfmathsetmacro{\x}{\x + \cubex + \offset}
\pgfmathsetmacro{\xx}{\xx/\divyz}
\draw[maxline,fill=maxfill] (\x+\xx/2,\cubey/2,\cubez/2) -- ++(-\cubex,0,0) -- ++(0,-\cubey,0) -- ++(\cubex,0,0) -- cycle;
\draw[maxline,fill=maxfill] (\x+\xx/2,\cubey/2,\cubez/2) -- ++(0-\xx,0,-\cubez) -- ++(0,-\cubey,0) -- ++(0+\xx,0,\cubez) -- cycle;
\draw[maxline,fill=maxfill] (\x+\xx/2,\cubey/2,\cubez/2) -- ++(-\cubex,0,0) -- ++(0-\xx,0,-\cubez) -- ++(\cubex,0,0) -- cycle;

\pgfmathsetmacro{\cubex}{\cubex*\multx}
\pgfmathsetmacro{\x}{\x + \cubex + \offset}
\draw[convline,fill=convfill] (\x+\xx/2,\cubey/2,\cubez/2) -- ++(-\cubex,0,0) -- ++(0,-\cubey,0) -- ++(\cubex,0,0) -- cycle;
\draw[convline,fill=convfill] (\x+\xx/2,\cubey/2,\cubez/2) -- ++(0-\xx,0,-\cubez) -- ++(0,-\cubey,0) -- ++(0+\xx,0,\cubez) -- cycle;
\draw[convline,fill=convfill] (\x+\xx/2,\cubey/2,\cubez/2) -- ++(-\cubex,0,0) -- ++(0-\xx,0,-\cubez) -- ++(\cubex,0,0) -- cycle;

\pgfmathsetmacro{\x}{\x + \cubex + \offset}
\draw[convline,fill=convfill] (\x+\xx/2,\cubey/2,\cubez/2) -- ++(-\cubex,0,0) -- ++(0,-\cubey,0) -- ++(\cubex,0,0) -- cycle;
\draw[convline,fill=convfill] (\x+\xx/2,\cubey/2,\cubez/2) -- ++(0-\xx,0,-\cubez) -- ++(0,-\cubey,0) -- ++(0+\xx,0,\cubez) -- cycle;
\draw[convline,fill=convfill] (\x+\xx/2,\cubey/2,\cubez/2) -- ++(-\cubex,0,0) -- ++(0-\xx,0,-\cubez) -- ++(\cubex,0,0) -- cycle;

\node[text width=3cm] at (\x+\xx/2-\xx+0.55,\cubey/2+0.12,\cubez/2-\cubez) {\scriptsize $64 \! \times \! 64 \! \times \! 32$};

\pgfmathsetmacro{\cubey}{\cubey/\divyz}
\pgfmathsetmacro{\cubez}{\cubez/\divyz}
\pgfmathsetmacro{\x}{\x + \cubex + \offset}
\pgfmathsetmacro{\xx}{\xx/\divyz}
\draw[maxline,fill=maxfill] (\x+\xx/2,\cubey/2,\cubez/2) -- ++(-\cubex,0,0) -- ++(0,-\cubey,0) -- ++(\cubex,0,0) -- cycle;
\draw[maxline,fill=maxfill] (\x+\xx/2,\cubey/2,\cubez/2) -- ++(0-\xx,0,-\cubez) -- ++(0,-\cubey,0) -- ++(0+\xx,0,\cubez) -- cycle;
\draw[maxline,fill=maxfill] (\x+\xx/2,\cubey/2,\cubez/2) -- ++(-\cubex,0,0) -- ++(0-\xx,0,-\cubez) -- ++(\cubex,0,0) -- cycle;

\pgfmathsetmacro{\cubex}{\cubex*\multx}
\pgfmathsetmacro{\x}{\x + \cubex + \offset}
\draw[convline,fill=convfill] (\x+\xx/2,\cubey/2,\cubez/2) -- ++(-\cubex,0,0) -- ++(0,-\cubey,0) -- ++(\cubex,0,0) -- cycle;
\draw[convline,fill=convfill] (\x+\xx/2,\cubey/2,\cubez/2) -- ++(0-\xx,0,-\cubez) -- ++(0,-\cubey,0) -- ++(0+\xx,0,\cubez) -- cycle;
\draw[convline,fill=convfill] (\x+\xx/2,\cubey/2,\cubez/2) -- ++(-\cubex,0,0) -- ++(0-\xx,0,-\cubez) -- ++(\cubex,0,0) -- cycle;

\pgfmathsetmacro{\x}{\x + \cubex + \offset}
\draw[convline,fill=convfill] (\x+\xx/2,\cubey/2,\cubez/2) -- ++(-\cubex,0,0) -- ++(0,-\cubey,0) -- ++(\cubex,0,0) -- cycle;
\draw[convline,fill=convfill] (\x+\xx/2,\cubey/2,\cubez/2) -- ++(0-\xx,0,-\cubez) -- ++(0,-\cubey,0) -- ++(0+\xx,0,\cubez) -- cycle;
\draw[convline,fill=convfill] (\x+\xx/2,\cubey/2,\cubez/2) -- ++(-\cubex,0,0) -- ++(0-\xx,0,-\cubez) -- ++(\cubex,0,0) -- cycle;

\pgfmathsetmacro{\x}{\x + \cubex + \offset}
\draw[convline,fill=convfill] (\x+\xx/2,\cubey/2,\cubez/2) -- ++(-\cubex,0,0) -- ++(0,-\cubey,0) -- ++(\cubex,0,0) -- cycle;
\draw[convline,fill=convfill] (\x+\xx/2,\cubey/2,\cubez/2) -- ++(0-\xx,0,-\cubez) -- ++(0,-\cubey,0) -- ++(0+\xx,0,\cubez) -- cycle;
\draw[convline,fill=convfill] (\x+\xx/2,\cubey/2,\cubez/2) -- ++(-\cubex,0,0) -- ++(0-\xx,0,-\cubez) -- ++(\cubex,0,0) -- cycle;

\node[text width=3cm] at (\x+\xx/2-\xx+0.23,\cubey/2+0.12,\cubez/2-\cubez) {\scriptsize $32 \! \times \! 32 \! \times \! 64$};

\pgfmathsetmacro{\cubey}{\cubey/2}
\pgfmathsetmacro{\cubez}{\cubez/2}
\pgfmathsetmacro{\x}{\x + \cubex + \offset}
\pgfmathsetmacro{\xx}{\xx/\divyz}
\draw[maxline,fill=maxfill] (\x+\xx/2,\cubey/2,\cubez/2) -- ++(-\cubex,0,0) -- ++(0,-\cubey,0) -- ++(\cubex,0,0) -- cycle;
\draw[maxline,fill=maxfill] (\x+\xx/2,\cubey/2,\cubez/2) -- ++(0-\xx,0,-\cubez) -- ++(0,-\cubey,0) -- ++(0+\xx,0,\cubez) -- cycle;
\draw[maxline,fill=maxfill] (\x+\xx/2,\cubey/2,\cubez/2) -- ++(-\cubex,0,0) -- ++(0-\xx,0,-\cubez) -- ++(\cubex,0,0) -- cycle;

\pgfmathsetmacro{\cubex}{\cubex*\multx}
\pgfmathsetmacro{\x}{\x + \cubex + \offset}
\draw[convline,fill=convfill] (\x+\xx/2,\cubey/2,\cubez/2) -- ++(-\cubex,0,0) -- ++(0,-\cubey,0) -- ++(\cubex,0,0) -- cycle;
\draw[convline,fill=convfill] (\x+\xx/2,\cubey/2,\cubez/2) -- ++(0-\xx,0,-\cubez) -- ++(0,-\cubey,0) -- ++(0+\xx,0,\cubez) -- cycle;
\draw[convline,fill=convfill] (\x+\xx/2,\cubey/2,\cubez/2) -- ++(-\cubex,0,0) -- ++(0-\xx,0,-\cubez) -- ++(\cubex,0,0) -- cycle;

\pgfmathsetmacro{\x}{\x + \cubex + \offset}
\draw[convline,fill=convfill] (\x+\xx/2,\cubey/2,\cubez/2) -- ++(-\cubex,0,0) -- ++(0,-\cubey,0) -- ++(\cubex,0,0) -- cycle;
\draw[convline,fill=convfill] (\x+\xx/2,\cubey/2,\cubez/2) -- ++(0-\xx,0,-\cubez) -- ++(0,-\cubey,0) -- ++(0+\xx,0,\cubez) -- cycle;
\draw[convline,fill=convfill] (\x+\xx/2,\cubey/2,\cubez/2) -- ++(-\cubex,0,0) -- ++(0-\xx,0,-\cubez) -- ++(\cubex,0,0) -- cycle;

\pgfmathsetmacro{\x}{\x + \cubex + \offset}
\draw[convline,fill=convfill] (\x+\xx/2,\cubey/2,\cubez/2) -- ++(-\cubex,0,0) -- ++(0,-\cubey,0) -- ++(\cubex,0,0) -- cycle;
\draw[convline,fill=convfill] (\x+\xx/2,\cubey/2,\cubez/2) -- ++(0-\xx,0,-\cubez) -- ++(0,-\cubey,0) -- ++(0+\xx,0,\cubez) -- cycle;
\draw[convline,fill=convfill] (\x+\xx/2,\cubey/2,\cubez/2) -- ++(-\cubex,0,0) -- ++(0-\xx,0,-\cubez) -- ++(\cubex,0,0) -- cycle;

\node[text width=3cm] at (\x+\xx/2-\xx+0.05,\cubey/2+0.12,\cubez/2-\cubez) {\scriptsize $16 \! \times \! 16 \! \times \! 128$};

\pgfmathsetmacro{\cubey}{\cubey/2}
\pgfmathsetmacro{\cubez}{\cubez/2}
\pgfmathsetmacro{\x}{\x + \cubex + \offset}
\pgfmathsetmacro{\xx}{\xx/\divyz}
\draw[maxline,fill=maxfill] (\x+\xx/2,\cubey/2,\cubez/2) -- ++(-\cubex,0,0) -- ++(0,-\cubey,0) -- ++(\cubex,0,0) -- cycle;
\draw[maxline,fill=maxfill] (\x+\xx/2,\cubey/2,\cubez/2) -- ++(0-\xx,0,-\cubez) -- ++(0,-\cubey,0) -- ++(0+\xx,0,\cubez) -- cycle;
\draw[maxline,fill=maxfill] (\x+\xx/2,\cubey/2,\cubez/2) -- ++(-\cubex,0,0) -- ++(0-\xx,0,-\cubez) -- ++(\cubex,0,0) -- cycle;

\pgfmathsetmacro{\x}{\x + \cubex + \offset}
\draw[convline,fill=convfill] (\x+\xx/2,\cubey/2,\cubez/2) -- ++(-\cubex,0,0) -- ++(0,-\cubey,0) -- ++(\cubex,0,0) -- cycle;
\draw[convline,fill=convfill] (\x+\xx/2,\cubey/2,\cubez/2) -- ++(0-\xx,0,-\cubez) -- ++(0,-\cubey,0) -- ++(0+\xx,0,\cubez) -- cycle;
\draw[convline,fill=convfill] (\x+\xx/2,\cubey/2,\cubez/2) -- ++(-\cubex,0,0) -- ++(0-\xx,0,-\cubez) -- ++(\cubex,0,0) -- cycle;

\pgfmathsetmacro{\x}{\x + \cubex + \offset}
\draw[convline,fill=convfill] (\x+\xx/2,\cubey/2,\cubez/2) -- ++(-\cubex,0,0) -- ++(0,-\cubey,0) -- ++(\cubex,0,0) -- cycle;
\draw[convline,fill=convfill] (\x+\xx/2,\cubey/2,\cubez/2) -- ++(0-\xx,0,-\cubez) -- ++(0,-\cubey,0) -- ++(0+\xx,0,\cubez) -- cycle;
\draw[convline,fill=convfill] (\x+\xx/2,\cubey/2,\cubez/2) -- ++(-\cubex,0,0) -- ++(0-\xx,0,-\cubez) -- ++(\cubex,0,0) -- cycle;

\pgfmathsetmacro{\x}{\x + \cubex + \offset}
\draw[convline,fill=convfill] (\x+\xx/2,\cubey/2,\cubez/2) -- ++(-\cubex,0,0) -- ++(0,-\cubey,0) -- ++(\cubex,0,0) -- cycle;
\draw[convline,fill=convfill] (\x+\xx/2,\cubey/2,\cubez/2) -- ++(0-\xx,0,-\cubez) -- ++(0,-\cubey,0) -- ++(0+\xx,0,\cubez) -- cycle;
\draw[convline,fill=convfill] (\x+\xx/2,\cubey/2,\cubez/2) -- ++(-\cubex,0,0) -- ++(0-\xx,0,-\cubez) -- ++(\cubex,0,0) -- cycle;

\node[text width=3cm] at (\x+\xx/2-\xx-0.05,\cubey/2+0.12,\cubez/2-\cubez) {\scriptsize $8 \! \times \! 8 \! \times \! 128$};




\pgfmathsetmacro{\cubex}{\cubex*2.5}
\pgfmathsetmacro{\cubey}{\cubey/2}
\pgfmathsetmacro{\cubez}{\cubez/2}
\pgfmathsetmacro{\x}{\x + \cubex + \offset}
\pgfmathsetmacro{\xx}{\xx/\divyz}
\draw[fcline,fill=fcfill] (\x+\xx/2,\cubey/2,\cubez/2) -- ++(-\cubex,0,0) -- ++(0,-\cubey,0) -- ++(\cubex,0,0) -- cycle;
\draw[fcline,fill=fcfill] (\x+\xx/2,\cubey/2,\cubez/2) -- ++(0-\xx,0,-\cubez) -- ++(0,-\cubey,0) -- ++(0+\xx,0,\cubez) -- cycle;
\draw[fcline,fill=fcfill] (\x+\xx/2,\cubey/2,\cubez/2) -- ++(-\cubex,0,0) -- ++(0-\xx,0,-\cubez) -- ++(\cubex,0,0) -- cycle;

\pgfmathsetmacro{\x}{\x + \cubex + \offset}
\draw[fcline,fill=fcfill] (\x+\xx/2,\cubey/2,\cubez/2) -- ++(-\cubex,0,0) -- ++(0,-\cubey,0) -- ++(\cubex,0,0) -- cycle;
\draw[fcline,fill=fcfill] (\x+\xx/2,\cubey/2,\cubez/2) -- ++(0-\xx,0,-\cubez) -- ++(0,-\cubey,0) -- ++(0+\xx,0,\cubez) -- cycle;
\draw[fcline,fill=fcfill] (\x+\xx/2,\cubey/2,\cubez/2) -- ++(-\cubex,0,0) -- ++(0-\xx,0,-\cubez) -- ++(\cubex,0,0) -- cycle;

\pgfmathsetmacro{\cubex}{\cubex/2}
\pgfmathsetmacro{\x}{\x + \cubex + \offset}
\draw[fcline,fill=fcdescmammo] (\x+\xx/2,\cubey/2,\cubez/2) -- ++(-\cubex,0,0) -- ++(0,-\cubey,0) -- ++(\cubex,0,0) -- cycle;
\draw[fcline,fill=fcdescmammo] (\x+\xx/2,\cubey/2,\cubez/2) -- ++(0-\xx,0,-\cubez) -- ++(0,-\cubey,0) -- ++(0+\xx,0,\cubez) -- cycle;
\draw[fcline,fill=fcdescmammo] (\x+\xx/2,\cubey/2,\cubez/2) -- ++(-\cubex,0,0) -- ++(0-\xx,0,-\cubez) -- ++(\cubex,0,0) -- cycle;

\node[text width=3cm] at (\x+\xx/2-\xx+0.6,\cubey/2+0.52,\cubez/2-\cubez) {\scriptsize $1 \! \times \! 1 \! \times \! 512$};

\draw[-stealth] (\x+\xx/2-\xx -\cubex/2,\cubey/2+0.4,0)--(\x+\xx/2-\xx -\cubex/2,\cubey/2 ,0);

\pgfmathsetmacro{\x}{\x + \cubex + \offset}
\draw[softmaxline,fill=softmaxfill] (\x+\xx/2,\cubey/2,\cubez/2) -- ++(-\cubex,0,0) -- ++(0,-\cubey,0) -- ++(\cubex,0,0) -- cycle;
\draw[softmaxline,fill=softmaxfill] (\x+\xx/2,\cubey/2,\cubez/2) -- ++(0-\xx,0,-\cubez) -- ++(0,-\cubey,0) -- ++(0+\xx,0,\cubez) -- cycle;
\draw[softmaxline,fill=softmaxfill] (\x+\xx/2,\cubey/2,\cubez/2) -- ++(-\cubex,0,0) -- ++(0-\xx,0,-\cubez) -- ++(\cubex,0,0) -- cycle;

\node[text width=3cm] at (\x+\xx/2-\xx+0.6,\cubey/2-0.72,\cubez/2-\cubez) {\scriptsize $1 \! \times \! 1 \! \times \! 2$};

\draw[-stealth] (\x+\xx/2-\xx -\cubex/2,\cubey/2-0.6 ,0)--(\x+\xx/2-\xx-\cubex/2,\cubey/2-0.2,0);

\pgfmathsetmacro{\x}{\x/2}
\pgfmathsetmacro{\y}{-1.2}
\pgfmathsetmacro{\cubex}{\cubex/1.5}
\pgfmathsetmacro{\cubey}{\cubey*3}
\pgfmathsetmacro{\cubez}{\cubez*3}

\draw[convline,fill=convfill] (\x+\xx/2,\y+\cubey/2,\cubez/2) -- ++(-\cubex,0,0) -- ++(0,-\cubey,0) -- ++(\cubex,0,0) -- cycle;
\draw[convline,fill=convfill] (\x+\xx/2,\y+\cubey/2,\cubez/2) -- ++(0-\xx,0,-\cubez) -- ++(0,-\cubey,0) -- ++(0+\xx,0,\cubez) -- cycle;
\draw[convline,fill=convfill] (\x+\xx/2,\y+\cubey/2,\cubez/2) -- ++(-\cubex,0,0) -- ++(0-\xx,0,-\cubez) -- ++(\cubex,0,0) -- cycle;

\node[text width=3cm] at (\x+1.7,\y+0.05,0) {\scriptsize convolution+ReLU};

\pgfmathsetmacro{\y}{\y-0.35}
\draw[maxline,fill=maxfill] (\x+\xx/2,\y+\cubey/2,\cubez/2) -- ++(-\cubex,0,0) -- ++(0,-\cubey,0) -- ++(\cubex,0,0) -- cycle;
\draw[maxline,fill=maxfill] (\x+\xx/2,\y+\cubey/2,\cubez/2) -- ++(0-\xx,0,-\cubez) -- ++(0,-\cubey,0) -- ++(0+\xx,0,\cubez) -- cycle;
\draw[maxline,fill=maxfill] (\x+\xx/2,\y+\cubey/2,\cubez/2) -- ++(-\cubex,0,0) -- ++(0-\xx,0,-\cubez) -- ++(\cubex,0,0) -- cycle;

\node[text width=3cm] at (\x+1.7,\y+0.05,0) {\scriptsize max pooling};

\pgfmathsetmacro{\y}{\y-0.35}
\draw[fcline,fill=fcfill] (\x+\xx/2,\y+\cubey/2,\cubez/2) -- ++(-\cubex,0,0) -- ++(0,-\cubey,0) -- ++(\cubex,0,0) -- cycle;
\draw[fcline,fill=fcfill] (\x+\xx/2,\y+\cubey/2,\cubez/2) -- ++(0-\xx,0,-\cubez) -- ++(0,-\cubey,0) -- ++(0+\xx,0,\cubez) -- cycle;
\draw[fcline,fill=fcfill] (\x+\xx/2,\y+\cubey/2,\cubez/2) -- ++(-\cubex,0,0) -- ++(0-\xx,0,-\cubez) -- ++(\cubex,0,0) -- cycle;

\node[text width=3cm] at (\x+1.7,\y+0.05,0) {\scriptsize fully connected};

\pgfmathsetmacro{\y}{\y-0.35}
\draw[softmaxline,fill=softmaxfill] (\x+\xx/2,\y+\cubey/2,\cubez/2) -- ++(-\cubex,0,0) -- ++(0,-\cubey,0) -- ++(\cubex,0,0) -- cycle;
\draw[softmaxline,fill=softmaxfill] (\x+\xx/2,\y+\cubey/2,\cubez/2) -- ++(0-\xx,0,-\cubez) -- ++(0,-\cubey,0) -- ++(0+\xx,0,\cubez) -- cycle;
\draw[softmaxline,fill=softmaxfill] (\x+\xx/2,\y+\cubey/2,\cubez/2) -- ++(-\cubex,0,0) -- ++(0-\xx,0,-\cubez) -- ++(\cubex,0,0) -- cycle;

\node[text width=3cm] at (\x+1.7,\y+0.05,0) {\scriptsize softmax};

\end{tikzpicture}
}
\caption{Proposed single modality basic CNN architecture. The input is a cropped lesion and the output is the softmax malignancy probability. A 512 dimensional vector (``descriptor'') is the last layer before the output layer.}
\label{fig:single_modality_net}
\end {figure}

We experimented with two different architectures: (1) Basic CNN with ReLU activation maps, max pooling and fully connected layers that was trained from scratch (Figure~\ref{fig:single_modality_net}); (2) GoogleNet~\cite{GoogleNet} previously trained over ImageNet.

\subsubsection{Multimodal network} 
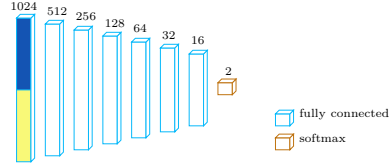
\begin {figure}[!t]
\centering
\resizebox {0.5\textwidth} {!} { %
\centering
\begin{tikzpicture}

\pgfmathsetmacro{\cubex}{0.3}
\pgfmathsetmacro{\cubey}{1}
\pgfmathsetmacro{\cubez}{0.2}

\pgfmathsetmacro{\multx}{1.5}
\pgfmathsetmacro{\divyz}{2}

\definecolor{convfill}{RGB}{255,255,255}
\definecolor{convline}{RGB}{0,0,0}

\definecolor{maxline}{RGB}{255,0,0}
\definecolor{maxfill}{RGB}{255,255,255}

\definecolor{fcfill}{RGB}{255,255,255}
\definecolor{fcline}{RGB}{6,185,251}
\definecolor{fcdescmammo}{RGB}{20,84,179}
\definecolor{fcdescus}{RGB}{249,249,120}

\definecolor{softmaxfill}{RGB}{255,255,255}
\definecolor{softmaxline}{RGB}{191,115,16}

\pgfmathsetmacro{\x}{0}
\pgfmathsetmacro{\offset}{0.6}

\pgfmathsetmacro{\xx}{0.5}

\draw[fcline,fill=fcdescus] (0,0,0) -- ++(-\cubex,0,0) -- ++(0,-3*\cubey,0) -- ++(\cubex,0,0) -- cycle;
\draw[fcline,fill=fcdescmammo] (0,0,0) -- ++(-\cubex,0,0) -- ++(0,-0.5*3*\cubey,0) -- ++(\cubex,0,0) -- cycle;
\draw[fcline,fill=convfill] (0,0,0) -- ++(0,0,-\cubez) -- ++(0,-3*\cubey,0) -- ++(0,0,\cubez) -- cycle;
\draw[fcline,fill=convfill] (0,0,0) -- ++(-\cubex,0,0) -- ++(0,0,-\cubez) -- ++(\cubex,0,0) -- cycle;

\node[text width=3cm] at (\x+\xx/0.6-\xx+0.7,\cubey/2-0.3,\cubez/2-\cubez) {\scriptsize $1024$};

\pgfmathsetmacro{\x}{\x + \offset}

\draw[fcline,fill=convfill] (\x,-0.125*\cubey,0) -- ++(-\cubex,0,0) -- ++(0,-2.75*\cubey,0) -- ++(\cubex,0,0) -- cycle;
\draw[fcline,fill=convfill] (\x,-0.125*\cubey,0) -- ++(0,0,-\cubez) -- ++(0,-2.75*\cubey,0) -- ++(0,0,\cubez) -- cycle;
\draw[fcline,fill=convfill] (\x,-0.125*\cubey,0) -- ++(-\cubex,0,0) -- ++(0,0,-\cubez) -- ++(\cubex,0,0) -- cycle;

\node[text width=3cm] at (\x+\xx/0.5-\xx+0.7,\cubey/2-0.4,\cubez/2-\cubez) {\scriptsize $512$};

\pgfmathsetmacro{\x}{\x + \offset}

\draw[fcline,fill=convfill] (\x,-0.25*\cubey,0) -- ++(-\cubex,0,0) -- ++(0,-2.5*\cubey,0) -- ++(\cubex,0,0) -- cycle;
\draw[fcline,fill=convfill] (\x,-0.25*\cubey,0) -- ++(0,0,-\cubez) -- ++(0,-2.5*\cubey,0) -- ++(0,0,\cubez) -- cycle;
\draw[fcline,fill=convfill] (\x,-0.25*\cubey,0) -- ++(-\cubex,0,0) -- ++(0,0,-\cubez) -- ++(\cubex,0,0) -- cycle;

\node[text width=3cm] at (\x+\xx/0.5-\xx+0.7,\cubey/2-0.5,\cubez/2-\cubez) {\scriptsize $256$};

\pgfmathsetmacro{\x}{\x + \offset}

\draw[fcline,fill=convfill] (\x,-0.375*\cubey,0) -- ++(-\cubex,0,0) -- ++(0,-2.25*\cubey,0) -- ++(\cubex,0,0) -- cycle;
\draw[fcline,fill=convfill] (\x,-0.375*\cubey,0) -- ++(0,0,-\cubez) -- ++(0,-2.25*\cubey,0) -- ++(0,0,\cubez) -- cycle;
\draw[fcline,fill=convfill] (\x,-0.375*\cubey,0) -- ++(-\cubex,0,0) -- ++(0,0,-\cubez) -- ++(\cubex,0,0) -- cycle;

\node[text width=3cm] at (\x+\xx/0.5-\xx+0.7,\cubey/2-0.65,\cubez/2-\cubez) {\scriptsize $128$};

\pgfmathsetmacro{\x}{\x + \offset}

\draw[fcline,fill=convfill] (\x,-0.5*\cubey,0) -- ++(-\cubex,0,0) -- ++(0,-2*\cubey,0) -- ++(\cubex,0,0) -- cycle;
\draw[fcline,fill=convfill] (\x,-0.5*\cubey,0) -- ++(0,0,-\cubez) -- ++(0,-2*\cubey,0) -- ++(0,0,\cubez) -- cycle;
\draw[fcline,fill=convfill] (\x,-0.5*\cubey,0) -- ++(-\cubex,0,0) -- ++(0,0,-\cubez) -- ++(\cubex,0,0) -- cycle;

\node[text width=3cm] at (\x+\xx/0.5-\xx+0.7,\cubey/2-0.75,\cubez/2-\cubez) {\scriptsize $64$};

\pgfmathsetmacro{\x}{\x + \offset}

\draw[fcline,fill=convfill] (\x,-0.625*\cubey,0) -- ++(-\cubex,0,0) -- ++(0,-1.75*\cubey,0) -- ++(\cubex,0,0) -- cycle;
\draw[fcline,fill=convfill] (\x,-0.625*\cubey,0) -- ++(0,0,-\cubez) -- ++(0,-1.75*\cubey,0) -- ++(0,0,\cubez) -- cycle;
\draw[fcline,fill=convfill] (\x,-0.625*\cubey,0) -- ++(-\cubex,0,0) -- ++(0,0,-\cubez) -- ++(\cubex,0,0) -- cycle;

\node[text width=3cm] at (\x+\xx/0.5-\xx+0.7,\cubey/2-0.9,\cubez/2-\cubez) {\scriptsize $32$};

\pgfmathsetmacro{\x}{\x + \offset}

\draw[fcline,fill=convfill] (\x,-0.75*\cubey,0) -- ++(-\cubex,0,0) -- ++(0,-1.5*\cubey,0) -- ++(\cubex,0,0) -- cycle;
\draw[fcline,fill=convfill] (\x,-0.75*\cubey,0) -- ++(0,0,-\cubez) -- ++(0,-1.5*\cubey,0) -- ++(0,0,\cubez) -- cycle;
\draw[fcline,fill=convfill] (\x,-0.75*\cubey,0) -- ++(-\cubex,0,0) -- ++(0,0,-\cubez) -- ++(\cubex,0,0) -- cycle;

\node[text width=3cm] at (\x+\xx/0.5-\xx+0.7,\cubey/2-1,\cubez/2-\cubez) {\scriptsize $16$};

\pgfmathsetmacro{\x}{\x + \offset}

\draw[softmaxline,fill=convfill] (\x,-1.35*\cubey,0) -- ++(-\cubex,0,0) -- ++(0,-0.25*\cubey,0) -- ++(\cubex,0,0) -- cycle;
\draw[softmaxline,fill=convfill] (\x,-1.35*\cubey,0) -- ++(0,0,-\cubez) -- ++(0,-0.25*\cubey,0) -- ++(0,0,\cubez) -- cycle;
\draw[softmaxline,fill=convfill] (\x,-1.35*\cubey,0) -- ++(-\cubex,0,0) -- ++(0,0,-\cubez) -- ++(\cubex,0,0) -- cycle;

\node[text width=3cm] at (\x+\xx/0.45-\xx+0.7,\cubey/2-1.65,\cubez/2-\cubez) {\scriptsize $2$};

\pgfmathsetmacro{\x}{\x + 2*\offset}

\draw[fcline,fill=convfill] (\x,-2,0) -- ++(-\cubex,0,0) -- ++(0,-0.25*\cubey,0) -- ++(\cubex,0,0) -- cycle;
\draw[fcline,fill=convfill] (\x,-2,0) -- ++(0,0,-\cubez) -- ++(0,-0.25*\cubey,0) -- ++(0,0,\cubez) -- cycle;
\draw[fcline,fill=convfill] (\x,-2,0) -- ++(-\cubex,0,0) -- ++(0,0,-\cubez) -- ++(\cubex,0,0) -- cycle;
\node[text width=3cm] at (\x+1.7,-2,0) {\scriptsize fully connected};  

\draw[softmaxline,fill=convfill] (\x,-2.5,0) -- ++(-\cubex,0,0) -- ++(0,-0.25*\cubey,0) -- ++(\cubex,0,0) -- cycle;
\draw[softmaxline,fill=convfill] (\x,-2.5,0) -- ++(0,0,-\cubez) -- ++(0,-0.25*\cubey,0) -- ++(0,0,\cubez) -- cycle;
\draw[softmaxline,fill=convfill] (\x,-2.5,0) -- ++(-\cubex,0,0) -- ++(0,0,-\cubez) -- ++(\cubex,0,0) -- cycle;
\node[text width=3cm] at (\x+1.7,-2.5,0) {\scriptsize softmax};

\end{tikzpicture}

}
\caption{Multimodality fully connected network architecture. The input is a concatenation of correspondence lesion descriptors extracted from mammography and ultrasound CNNs. The output is the softmax malignancy probability.}
\label{fig:multimodal_net}
\end {figure}

Figure~\ref{fig:multimodal_net} presents the multimodal fully connected network, consisting of 7 layers. High-level perceptual descriptors of matched lesions were extracted from both trained single-modality networks and combined by concatenation. The concatenated vector is then used as an input for the multimodal network, which eventually provides the final malignancy probability of the input lesion.

\subsection{Implementation Details}
\label{sec:implementation_details}

\subsubsection{Loss function}
Both single and multi modal classifiers were trained using the same loss function in each experiment. To enrich diversity, we experimented with two loss functions: (1) BCE - Binary Cross Entropy loss; (2) LMCL - Large Margin Cosine Loss~\cite{CosFace} which is commonly used in Face Recognition tasks. LMCL defines a decision margin in the cosine space and learns discriminative features by maximizing inter-class and minimizing intra-class cosine margin.

\subsubsection{Training method}
\begin{figure}[!t]
\centering
\includegraphics[width=0.25\textwidth]{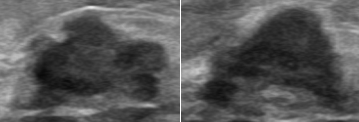}
\caption{Same lesion captured in different views in ultrasound screening.}
\label{fig:us_variance}
\end{figure}

Lesion patches from different modalities can be utilized for classification in different manners. Unfortunately, image registration is almost impossible because of the difference in mammography and ultrasound imaging techniques. Moreover, even ultrasound images of the same lesion are highly different, as the images are captured in various views and the breast is easily deformed by the mechanical pressure applied by the transducer (see Figure~\ref{fig:us_variance}).

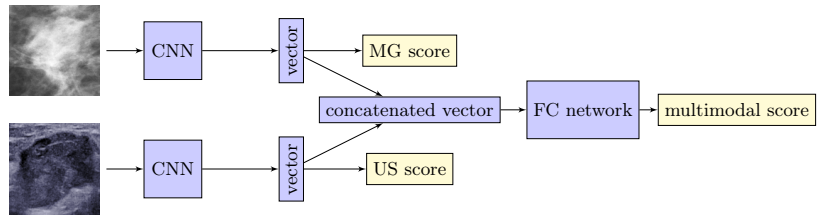
\begin {figure}[!t]
\centering
\resizebox {0.9\textwidth} {!} { %
\centering
%
%
%
\begin{tikzpicture}[node distance=2cm,auto,>=latex']



%
\tikzstyle{int}=[draw, fill=blue!20, minimum size=3em]
\tikzstyle{vector}=[draw, fill=blue!20, minimum size=1em]
\tikzstyle{score}=[draw, fill=yellow!20, minimum size=15pt]

\node (a) {\includegraphics[width=.125\textwidth]{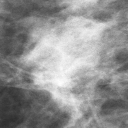}};
\node [int] (b) [right of=a] {CNN};
\node [vector] (c) [right of=b, rotate=90]{vector};
\node[score] (d) [right of=c]{MG score};

\node (a1) [below of=a] {\includegraphics[width=.125\textwidth]{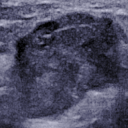}};
\node [int] (b1) [right of=a1] {CNN};
\node [vector] (c1) [right of=b1, rotate=90]{vector};
\node[score] (d1) [right of=c1]{US score};

\node[vector] (a2) at ($(d)!0.5!(d1)$) {concatenated vector};
\node[int] (b2) [right of=a2, node distance=9em] {FC network};
\node[score, node distance=8em] (c2) [right of=b2]{multimodal score};

\path[->] (a) edge node {} (b);
\path[->] (b) edge node {} (c);
\draw[->] (c) edge node {} (d) ;
\draw[->] (c) edge node {}
(a2) ;

\path[->] (a1) edge node {} (b1);
\path[->] (b1) edge node {} (c1);
\draw[->] (c1) edge node {} (d1) ;
\draw[->] (c1) edge node {}
(a2) ;

\path[->] (a2) edge node {} (b2);
\path[->] (b2) edge node {} (c2);
\end{tikzpicture}
}
\caption{End-to-end training method: all three classifiers are trained at the same time, while the loss is the sum of all three losses.}
\label{fig:end2end}
\end {figure}

Therefore, we make use of the coupled mammography-ultrasound lesions by combining them in the feature space instead of image space. For completeness, we show two different training methods: (1) We first train each single modality network separately, then combine high-level feature data from both networks and feed it as an input for training the multimodal network; (2) End-to-end training, illustrated in Figure~\ref{fig:end2end}, in which we train all three networks  (two single modality networks and one fully connected after feature combination) concurrently. The loss function is the sum of the losses from each of the three networks. In this approach, the performance of each network is tied to the other two's.

\section{Experiments}

\subsection{Leave-one-out}
Given 153 mammography-ultrasound lesion pairs, we randomly selected 120 fixed pairs for the leave-one-out experiments, benign and malignant being equally distributed. The remaining 33 lesions were held out as validation set for hyper-parameter tuning. 

We ran 8 different experiments, one for each combination of previously mentioned configurations: training method, model architecture and loss function. All the experiments were performed using
the ``leave-one-out'' methodology. In each phase, 119 out of 120 lesions were used for training and a single lesion, different in every phase, was used for testing. Finally, the test lesion obtained three scores, one for each modality and one combined, representing the average malignancy probabilities of all its appearances in the dataset.

Results were evaluated by means of AUC (area under the ROC curve), calculated from all test scores. As can be seen in Table~\ref{table:AUC}, combining mammography and ultrasound features improved results in most experiments. In fact, only in one experiment results deteriorated due to the combination of modalities. 
Clearly, using transfer learning outperforms training from scratch, likely because of the small dataset size. Moreover,
two steps of training (each modality first and combined descriptors afterwards) achieve better results than an end-to-end training. No significant difference is observed in the performance of the tested loss functions.

\subsubsection{Comparison to state-of-the-art models} Our method is based on combining mammography with ultrasound. However, as previous authors haven't discussed their exact models' design~\cite{MG_US_ensemble_classification}, we report the results of two baselines, each using a single modality, and compare them to our single modality networks. As each model was trained over a different dataset, it may be confusing and even meaningless to directly compare reported results. Therefore, for qualitative assessment of our model, we trained these models over our own dataset.

\paragraph{Mammography} The patch-level network proposed by~\cite{NYU_mammo} was trained on our mammography dataset. Based on DenseNet121~\cite{DenseNet} and transfer learning, it achieved an AUC of 0.86 - better than our trained from scratch CNN (0.76), but inferior to GoogleNet (0.89).

\paragraph{Ultrasound} Training VGG16 model~\cite{VGG16} previously trained over ImageNet, on our ultrasound dataset, as suggested by~\cite{US_classification_reference}, yielded AUC of 0.81. It is worth mentioning that the reported AUC on the original dataset of Hijab et al. was much higher (0.97), which may suggest that their method is sensitive to the specific training data used. The obtained AUC is better than our trained from scratch CNN (0.75), but inferior to GoogleNet (0.88).

\begin{table}
\setlength{\tabcolsep}{6pt}
\caption{AUC results of all experiments, reported on test set. Scores order is as follows: mammography/ultrasound/combined.}\label{tab1}
\begin{tabular}{l l c c}
\toprule
\textbf{Training} & \textbf{Loss} & \multicolumn{2}{c}{\textbf{Initialization}} \\
\textbf{Method} & \textbf{Function} & \textbf{\scriptsize From scratch (CNN)} & \textbf{\scriptsize Transfer Learning (GoogleNet)}\\
\toprule
Separate & BCE & {0.76/0.75/\bfseries0.79} & {0.89/0.88/\bfseries0.94}\\
\cline{2-4}
 & LMCL & {0.73/0.79/\bfseries0.82} & {0.88/0.87/\bfseries0.89}\\
\hline
End-to-end & BCE & {0.74/0.78/\bfseries0.79} & {\begin{bfseries}0.82\end{bfseries}/0.81/\bfseries0.82}\\
\cline{2-4}
 & LMCL & {0.72/0.78/\bfseries0.79} & {\begin{bfseries}0.84\end{bfseries}/0.80/0.78}\\
\bottomrule
\end{tabular}
\label{table:AUC}
\end{table}

\subsection{Reader study}
\begin{figure}[!t]
\centering
\includegraphics[width=0.55\textwidth]{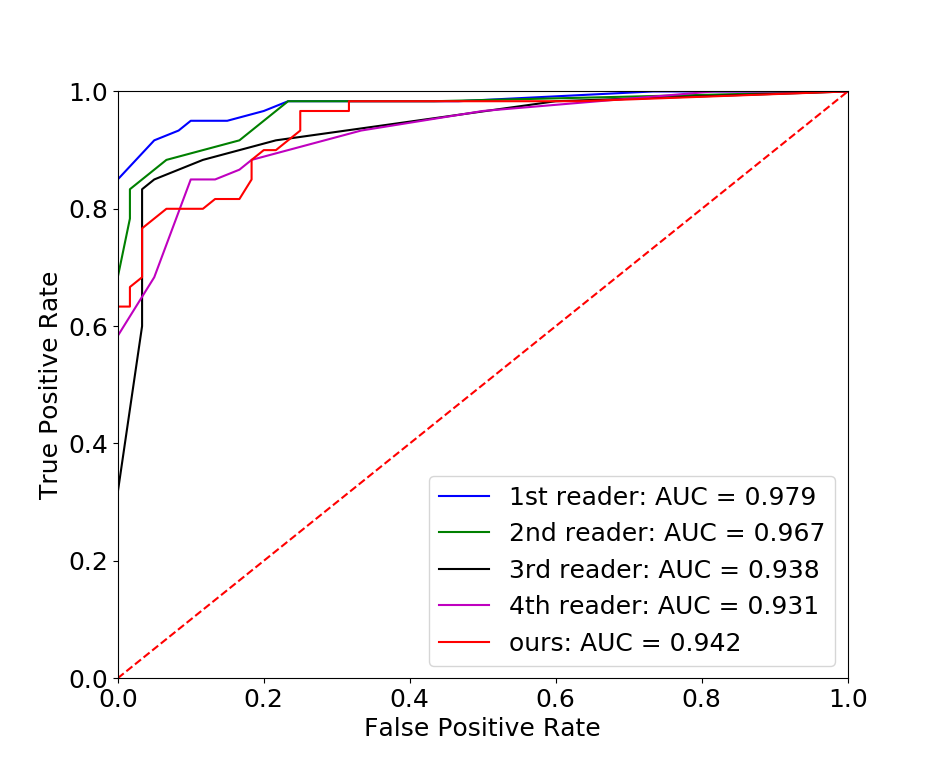}
\caption{ROC curves of our model and each reader.}
\label{fig:readers_study}
\end{figure}

To compare the proposed method with human radiologists, we performed a simplified reader study with 4 experienced radiologists. 120 pairs of corresponding mammography-ultrasound lesion images, taken from the biopsy-proven leave-one-out experiment dataset, were visually assigned a malignancy rate (from 0 to 10) by each of the participating radiologists separately.
The AUCs achieved by the readers were: 0.931, 0.938, 0.967 and 0.979, compared to 0.942 for our best model. ROC curves are shown in figure~\ref{fig:readers_study}. These results suggest that the proposed model performed similarly to an average radiologist. 

\subsection{Model insight}
Is the model paying attention to the same attributes as radiologists when predicting whether a lesion is malignant? To gain insight about this question, we applied the Grad-CAM algorithm~\cite{GradCAM} to our best trained GoogleNet. Grad-CAM produces a gradient-based heat-map that highlights input parts that most influenced the output prediction.

\begin{figure}[!t]
\centering
\includegraphics[width=0.6\textwidth]{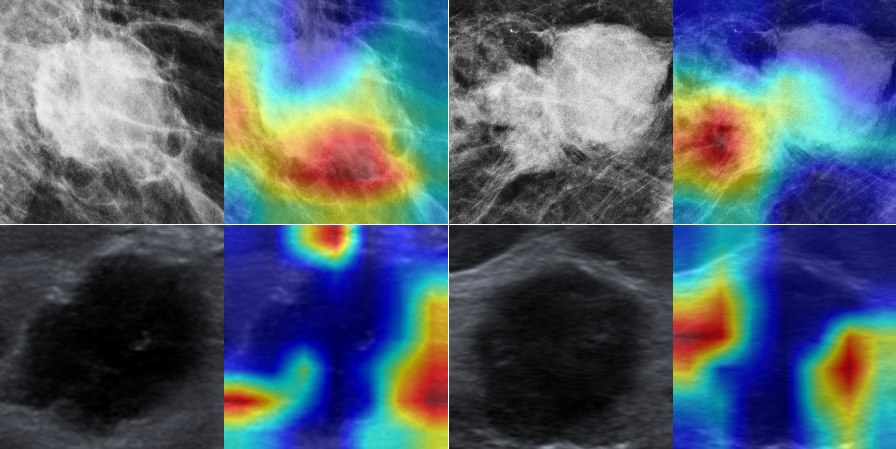}
\caption{Examples of malignant lesions from the training set with their Grad-CAM visualizations (Top-mammography, Bottom-ultrasound).}
\label{fig:GradCAM}
\end{figure}

In Figure~\ref{fig:GradCAM}, we present several mammography and ultrasound malignant examples from the training set, with their Grad-CAM computations. ``Hotter'' areas indicate attended regions. We observe that for malignant lesions, the model appears to rely significantly (hot colors in heat map)  on  the lesion boundaries,  especially where irregular features are encountered, in agreement with the radiologist diagnostic methodology.

\section{Conclusion}
We propose a deep-learning method for the classification of breast lesions that combines mammography and ultrasound input images. We show that by combining high-level perceptual features from both modalities, the classification performance is improved. Furthermore, the proposed method is shown to perform similarly to an average radiologist, surpassing two out of four radiologists participating in a reader study. The promising results suggest the proposed method may become a valuable decision support tool for multimodal classification of breast lesions. In future research, further validation on a larger dataset should be performed. The proposed method may be generalized by incorporating additional imaging modalities, such as breast MRI, as well as medical background information of the patient~\cite{text_mammo_pathology}.

%

%
%
%
%

\end{document}